\documentclass[rnote]{aa} 
\usepackage{graphicx}
\usepackage{txfonts}
\usepackage{natbib}
\bibpunct{(}{)}{;}{a}{}{,}

\begin{document}
   \title{The HI absorption distance of HESS J1943+213 favours its extragalactic nature}

\author{D.A. Leahy \inst{1}
\and
     W.W. Tian  \inst{2}}
\authorrunning{Leahy \& Tian}
\titlerunning{HESS J1943+213 is an Extragalactic Source}
\offprints{tww@bao.ac.cn}
\institute{Department of Physics \& Astronomy, University of Calgary, Calgary, Alberta T2N 1N4, Canada;\\
\and
National Astronomical Observatories, CAS, Beijing 100012, China.  tww@bao.ac.cn}
\date{Received  2011/Accepted 25th, Jan. 2012}

 
\abstract
   {The H.E.S.S. collaboration (Abramowski et al. 2011) dicovered a new TeV point-like source HESS J1943+213 in the Galactic plane and suggested three possible low-energy-band counterparts: a $\gamma$-ray binary, a pulsar wind nebula (PWN), or a BL Lacertae object.} 
{We measure the distance to the radio counterpart G57.76-1.29 of HESS J1943+213.}
{We analyze Very Large Array observations to obtain a reliable HI absorption spectrum.}
 {The resulting distance limit is $\ge$ 16 kpc.
This distance 
strongly supports that HESS J1943+213 is an extragalactic source, consistent with the preferred counterpart of the HESS collaboration}.
\maketitle

\section{Introduction}
Very-high-energy (VHE) $\gamma$-ray observations shed a light on the question of origin and acceleration of cosmic rays (Caprioli 2011). 
A multi-wavelength approach helps to solve the question, especially by seeking VHE sources' counterparts in lower energy bands. 
More than 110 VHE $\gamma$-ray sources have been detected by ground-based high energy telescopes recently (http://www.mppmu.mpg.de/~rwagner/sources/).  About 30 of them have no counterparts identifed so far. 
The identification of counterparts plays a key role in understanding the nature of the VHE sources and in distinguishing different emission mechanisms for TeV $\gamma$-rays (Halpern \& Gotthelf 2010, Misanovic et al. 2011, Tian et al. 2008). 

HESS J1943+213 was recently discovered by the H.E.S.S. collaboration (Abramowski et al. 2011)
and was suggested to have three possible low-energy-band counterparts: a $\gamma$-ray binary, a pulsar wind nebula (PWN), or a BL Lacertae object. HESS J1943+213 has a compact radio counterpart G57.76-1.29 which has been observed in the VGPS (also a NVSS source, NVSS J194356+211826, see Condon et al. 1998). 
 
We analyze 1.4 GHz continuum and HI-line observations of G57.76-1.29 from Very Large Array (VLA) Galactic Plane Survey (VGPS, Stil et al. 2006).
Based on our well-tested methods to build HI absorption spectrum against a continuum background radio source, we extract an HI absorption spectrum towards G57.76-1.29/HESS J1943+213. We report our results in this research note. 

\section{Results and discussion}
\label{res}

\begin{figure}
\vspace{50mm}
\begin{picture}(50,50)
\put(-40,230){\includegraphics{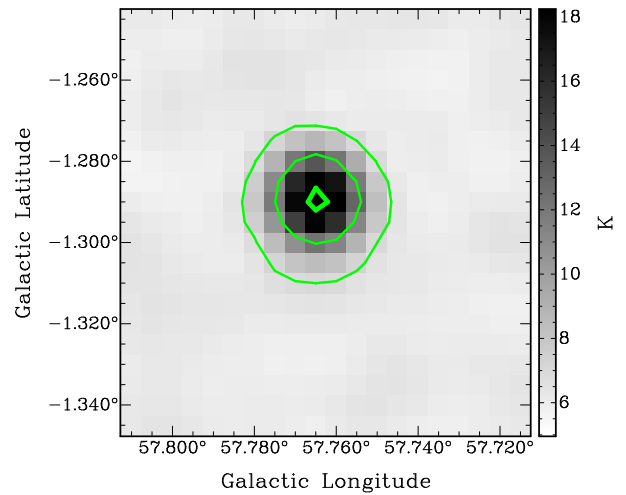}}
\end{picture}

   \caption{1.4-GHz VLA image of HESS J1943+213 (contours 7, 15, 20 K) from the VGPS data.}
   \end{figure}

\begin{figure}
   \centering
   \includegraphics*[width=8cm]{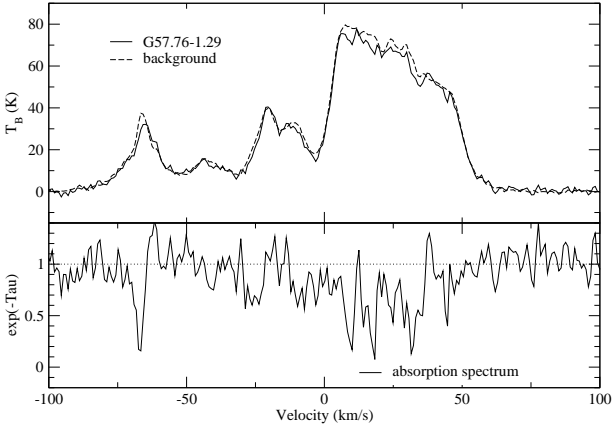}
      \caption{Top panel: HI emission spectrum towards J1943+2118. Bottom panel: HI absorption spectrum. The standard deviation of exp(-Tau) is 0.14, taken from the velocity range of 60 to 100 km/s (the mean value is 1.04)} 
   \end{figure}

\begin{figure}
   \vspace{50mm}
\begin{picture}(50,50)
\put(-40,230){\includegraphics{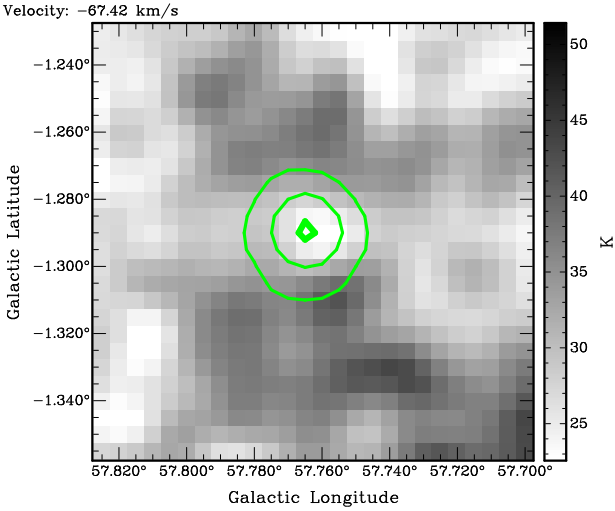}}
\end{picture}
\caption{The HI channel map at -67 km s$^{-1}$ with 1420 MHz continuum contours (the same as in Figure 1).}
   \end{figure}

Fig. 1 shows the 1420 MHz continuum map in a 0.1$^\circ$ field containing G57.76-1.29.
The VGPS data has an angular resolution of $1\arcmin$ and an rms noise inthe HI Line of 2 K (Stil et al. 2006).
It is a good database to study Galactic supernova remnants (SNRs) which usually extend from a
few arcmin to $\sim$1 degree in size. 
We have previously used the data to study some interesting SNRs (see a review paper by Leahy \& Tian 2010) 
and have also developed methods to estimate kinemtic distances to them by analyzing 21cm continuum, HI-line and CO-line data (Tian, Leahy \& Wang 2007; Leahy \& Tian 2008). 

Fig. 2 displays the HI emission spectra of source and background regions (top panel) and the absorption spectrum (bottom panel) towards G57.76-1.29. The background region was taken as
directly surrounding the source region. 
Pixels were assigned to source or background regions depending on whether they have 1420 MHz continuum brightness above or below a specified threshold, respectively. 
We used the program meanlev in the DRAO (Dominion Radio Astrophysical Observatory) Export Software Package to calculate the source and background spectra.
Then we used our methods, which take into account the background continuum radiation, to calculate optical depth in the form $exp(-\tau)$.
The resulting HI absorption spectrum shows absorption at all Galactic velocities. 
This includes absorption at positive velocities, from the interstellar medium interior to the solar circle, up to the tangient point velocity. 
Also we see absorption for negative velocities above $\sim$ -67 km/s, which corresponds approximately to the outer edge of the Galaxy.
The lowest velocity absorption feature (Fig. 2) appears at $\sim$ -67 km s$^{-1}$ and is associated with the emission peak from gas in the outer galaxy. 
This absorption feature's reality is further supported by the continuum-subtracted HI channel map at the peak of the absorption at -67 km s$^{-1}$ (Fig. 3). 
This clearly shows the spatial association between the lower 
HI brightness temperature, caused by absorption, and the peak of the 1420 MHz continuum source. 
We estimate the uncertainties in our HI absorption spectrum using the standard derivation of
exp(-$\tau$), taken from the velocity range 60 to 100 km/s, where the spectrum is dominated by noise.  The resulting  standard derivation is 0.14.
This yields that the absorption feature at $\sim$ -67 km s$^{-1}$ is $\sim$6 times the noise level and is real.  

The velocity limit gives a lower kinematic distance limit of $\sim$ 16 kpc to G57.76-1.29, taking a standard Galactic circular rotation model (V$_o$=220 km s$^{-1}$, R$_o$=8.5 kpc).
Our absorption spectrum is consistent and gives a picture of absorption out to the most negative velocities for which there is emission, so that G57.76-1.29 is an extragalactic source. 

In summary, HESS J1943+213 has three possible low-energy-band counterparts (Abramowski et al. 2011): a $\gamma$-ray binary, a pulsar wind nebula (PWN), or a BL Lacertae object.
Our observations and analysis show that HESS J1943+213, and its confirmed radio counterpart G57.76-1.29, is an extragalactic source.
  
\begin{acknowledgements}
DAL acknowledges support from the Natural Sciences and Engineering Research Council of Canada. WWT receives supports from the NSFC (011241001), BeiRen program
of the CAS (034031001) and State Key Development Program for Basic Research of China Ministry of Science and Technology (2012CB821800).
\end{acknowledgements}
\bibliographystyle{aa}
\bibliography{ref}

\begin{thebibliography}{28}
\expandafter\ifx\csname natexlab\endcsname\relax\def\natexlab#1{#1}\fi

\bibitem[{{Abramowski} {et~al.}(2011){Abramowski}, {Acero}, {Aharonian},
  {Akhperjanian}, {Anton}, {Balzer}, {Barnacka}, {Barres de Almeida},
  {Bazer-Bachi}, {Becherini}, {Becker}, {Behera}, {Bernl{\"o}hr}, {Bochow},
  {Boisson}, {Bolmont}, {Bordas}, {Borrel}, {Brucker}, {Brun}, {Brun}, {Bulik},
  {B{\"u}sching}, {Carrigan}, {Casanova}, {Cerruti}, {Chadwick}, {Charbonnier},
  {Chaves}, {Cheesebrough}, {Chounet}, {Clapson}, {Coignet}, {Colom}, {Conrad},
  {Dalton}, {Daniel}, {Davids}, {Degrange}, {Deil}, {Dickinson},
  {Djannati-Ata{\"i}}, {Domainko}, {Drury}, {Dubois}, {Dubus}, {Dyks}, {Dyrda},
  {Egberts}, {Eger}, {Espigat}, {Fallon}, {Farnier}, {Fegan}, {Feinstein},
  {Fernandes}, {Fiasson}, {Fontaine}, {F{\"o}rster}, {F{\"u}{\ss}ling},
  {Gallant}, {Gast}, {G{\'e}rard}, {Gerbig}, {Giebels}, {Glicenstein},
  {Gl{\"u}ck}, {Goret}, {G{\"o}ring}, {H{\"a}ffner}, {Hague}, {Hampf},
  {Hauser}, {Heinz}, {Heinzelmann}, {Henri}, {Hermann}, {Hinton}, {Hoffmann},
  {Hofmann}, {Hofverberg}, {Holler}, {Horns}, {Jacholkowska}, {de Jager},
  {Jahn}, {Jamrozy}, {Jung}, {Kastendieck}, {Katarzy{\'n}ski}, {Katz},
  {Kaufmann}, {Keogh}, {Khangulyan}, {Kh{\'e}lifi}, {Klochkov}, {Klu{\'z}niak},
  {Kneiske}, {Komin}, {Kosack}, {Kossakowski}, {Laffon}, {Lamanna}, {Lennarz},
  {Lohse}, {Lopatin}, {Lu}, {Marandon}, {Marcowith}, {Masbou}, {Maurin},
  {Maxted}, {McComb}, {Medina}, {M{\'e}hault}, {Nguyen}, {Moderski}, {Moulin},
  {Naumann}, {Naumann-Godo}, {de Naurois}, {Nedbal}, {Nekrassov}, {Nicholas},
  {Niemiec}, {Nolan}, {Ohm}, {Olive}, {de O{\~n}a Wilhelmi}, {Opitz},
  {Ostrowski}, {Panter}, {Paz Arribas}, {Pedaletti}, {Pelletier}, {Petrucci},
  {Pita}, {P{\"u}hlhofer}, {Punch}, {Quirrenbach}, {Raue}, {Rayner}, {Reimer},
  {Reimer}, {Renaud}, {de los Reyes}, {Rieger}, {Ripken}, {Rob}, {Rosier-Lees},
  {Rowell}, {Rudak}, {Rulten}, {Ruppel}, {Ryde}, {Sahakian}, {Santangelo},
  {Schlickeiser}, {Sch{\"o}ck}, {Sch{\"o}nwald}, {Schulz}, {Schwanke},
  {Schwarzburg}, {Schwemmer}, {Shalchi}, {Sikora}, {Skilton}, {Sol},
  {Spengler}, {Stawarz}, {Steenkamp}, {Stegmann}, {Stinzing}, {Stycz},
  {Sushch}, {Szostek}, {Tavernet}, {Terrier}, {Tibolla}, {Tluczykont},
  {Valerius}, {van Eldik}, {Vasileiadis}, {Venter}, {Vialle}, {Viana},
  {Vincent}, {Vivier}, {V{\"o}lk}, {Volpe}, {Vorobiov}, {Vorster}, {Wagner},
  {Ward}, {Wierzcholska}, {Zajczyk}, {Zdziarski}, {Zech}, {Zechlin}, {Burnett},
  \& {Hill}}]{hess_disc}
{Abramowski}, A., {Acero}, F., {Aharonian}, F., {et~al.} 2011, \aap, 529, A49

\bibitem[{{Caprioli(2011)}}]{review}
{Caprioli}, D. 2011, JCAP, 5, 26


\bibitem[{{Condon} {et~al.}(1998){Condon}, {Cotton}, {Greisen}, {Yin},
  {Perley}, {Taylor}, \& {Broderick}}]{nvss}
{Condon}, J.~J., {Cotton}, W.~D., {Greisen}, E.~W., {et~al.} 1998, \aj, 115,
  1693

\bibitem[{{Halpern} \& {Gotthelf}(2010)}]{HESS J1713-381}
{Halpern}, J.P., {Gotthelf}, E.V., 2010, \apj, 725, 1384 



\bibitem[{{Leahy} \& {Tian} (2008)}]{Kes 75}
{leahy}, D.A., {Tian}, W.W., 2008, A\&A, 408, L25

\bibitem[{{Leahy} \& {Tian} (2010)}]{review} {leahy}, D.A., {Tian}, W.W., 2010,
ASPC, 438, 365

\bibitem[{{Misanovic},{Kargaltsev},{Pavlov}(2011)}]{W41} {Misanovic}, Z., {Kargaltsev}, O., {Pavlov}, G.G. 2011, \apj, 735, 33

\bibitem[{{Stil} {et~al.}(2006){Stil}, {Taylor}, {Dickey}, {Kavars}, {Martin},
  {Rothwell}, {Boothroyd}, {Lockman}, \& {McClure-Griffiths}}]{VLAGPS}
{Stil}, J.~M., {Taylor}, A.~R., {Dickey}, J.~M., {et~al.} 2006, \aj, 132, 1158

\bibitem[{{Tian}{et~al.}(2007){Tian}, {Leahy}, {Wang}}]{G18.8}
{Tian}, W.W., {Leahy}, D.A., {Wang}, Q.D. 2007, A\&A, 474, 541

\bibitem[{{Tian}{et~al.}(2008){Tian}, {Leahy}, {Haverkorn}, {Jiang}}]{G353.6-0.7}
{Tian}, W.W., {Leahy}, D.A., {Haverkorn}, M., {Jiang}, B. 2008, \apj, 679, L85


\end{thebibliography}

\end{document}